\documentclass[%
 reprint,
 amsmath,amssymb,
 aps,
floatfix,
]{revtex4-2}

\usepackage{graphicx}
\usepackage{dcolumn}
\usepackage{bm}
\usepackage{subfigure}
\usepackage{wrapfig}
\usepackage{hyperref}

\begin{document}

\preprint{APS/123-QED}

\title{Optimizations of Multilevel Quantum  Engine
with \textit{N} Noninteracting Fermions Based on Lenoir Cycle}

\author{Ade Fahriza$^1$}
\author{Trengginas E P Sutantyo$^1$}
\email{trengginasekaputra@sci.unand.ac.id}
\author{Zulfi Abdullah$^1$}
\affiliation{%
$^1$Nuclear and Theoretical Physics Laboratory, Department of Physics, Faculty of Mathematics and Natural Science, Andalas University, Indonesia\\}%
\date{\today}

\begin{abstract}
We consider optimizations of Lenoir engine within a quantum dynamical field consisting of $N$ noninteracting fermions trapped in multilevel infinite potential square-well. Fermions play role as working substance of the engine with each particle nested at different level of energy. We optimized this quantum heat engine model by analysing the physical parameter and deriving the optimum properties of the engine model. The model we investigated consists of one high-energy heat bath and one low-energy sink bath. Heat leakage occurs between these two bathes as expected will degenerate the efficiency of quantum heat engine model. The degeneration increased as we raised the constant parameter of heat leakage. We also obtained loop curves in dimensionless power vs. efficiency of the engine, which efficiency is explicitly affected by heat leakage, but in contrast for the power output. From the curves, we assured that the efficiency of the engine would go back to zero as we raised compression ratio of engine with leakage. Lastly, we checked Clausius relations for each model with various levels of heat leakage. We found that models with leakage have a reversible process on specific compression ratios for each variation of heat leakage. Nevertheless, the compression ratio has limitations because of the $\oint dQ/E>0$ after the reversible point, i.e. violates the Clausius relation. 
\end{abstract}

\maketitle


\section{Introduction}
Heat engine in quantum dynamical field study has been pioneered by Scovil and Schulz-DuBois \cite{HEDScovil.1959}. Replacing classical view to quantum view of thermodynamic gives significant impact to knowledge of heat engine models, such as solution for the fundamental restriction of thermodynamics in macro physics \cite{AMani.2019, KP.2018, RUzdin.2015, HTQuan.2007, TDKieu.2006, MOScully.2001}, as well as the study of micro control of adiabatic or particles behaviour in heat engine models \cite{GHXu.2021, XShi.2021, LLi.2021, BCakmak.2019, OAbah.2019, OAbah.2018, GThomas.2017}. Moreover, in quantum models, heat engine with adiabatic or isentropic process which require quasi-static process can be done in finite time. This evolution of view, Classical to Quantum, is approved to be realized by the result of analogues in Bender model \cite{CMBender.2002, CMBender.2000}.

Quantum model of heat engines has wide probability to be investigated, e.g. considering the particle types as working substance \cite{FAbdillah.2021, MMAli.2020, MSAkbar.2018, EMunoz.2012}, the quantum system used to construct the quantum heat engine \cite{JUm.2021, JLDdeOliveira.2021, MFAnka.2021, Papadatos.2021, TEPSutantyo.2020, JJFernandez.2019, BKAgarwalla.2017, IHBelfaqih.2015, RWang.2012}, the thermodynamic cycle used in model engine \cite{FAltintas.2019, DPSetyo.2018, SSeah.2018, YYin.2017, DGupta.2017, TEPSutantyo.2015, ELatifah.2013}, etc. The three-steps thermodynamic process of Lenoir heat engine has been tried to be optimized model for future engine \cite{RWang.2021, MHAhmadi.2019, DPGeorgiou.2000}. Extracting an optimal model from various combination idea of quantum heat engine is what make this field of study interesting \cite{CTang.2021, SSingh1.2020, TMMendonca.2020, EA.2018, GWatanabe.2017, OAbah.2012, YGe.2009, YRezek.2006}.

Saputra \cite{YDSaputra.2021} considered a single particle as working substance in a one-dimensional box of quantum Lenoir heat engine model. They obtained that the efficiency of the quantum heat engine can be higher than the classical at same compression ratio if specific heat value of quantum heat engine is greater than the classical. The result proved Lenoir cycle can be applied to quantum field study.

Singh \cite{SSingh2.2020} presented a quantum Brayton heat engine with noninteracting fermions in one dimensional box model. The efficiency and irreversibility of engine have not affected by the number of the fermions in system. However, the amount of power product of the model is increased as the number of fermions in the system is also increased. 

Wang and He \cite{JWang.2012} studied one dimensional quantum heat engine based on the Carnot cycle considering the heat leakage between two energy baths. They used noninteracting fermion particles as working substance and obtained loop shapes for dimensionless power vs. efficiency curves which means the existence of heat leakage in heat engine makes the efficiency of the engine back to zero point as the volume ratio increase. The parameter of heat leakage, $\alpha$, highly impacts the efficiency value obtained, but not for the power output of the heat engine. 

In this study, we would investigate the Lenoir heat engine’s model in quantum thermodynamic view which heat leakage applied to the engine. Considering the leakage, this model would provide physical knowledge for quantum heat engine realization, i.e. representing the interaction with the environment. This model would be operated with N-number of noninteracting fermions within multilevel quantum system as the working substance. We would also try to derive the optimized models of this engine as a reference for quantum Lenoir heat engine in further development. In the last, we also inspect the irreversibility of the engine through Clausius inequality.

\section{Formalism of Quantum Lenoir Engine}\label{sec2} 

\begin{figure}[htp]
\includegraphics[width=0.5\textwidth]{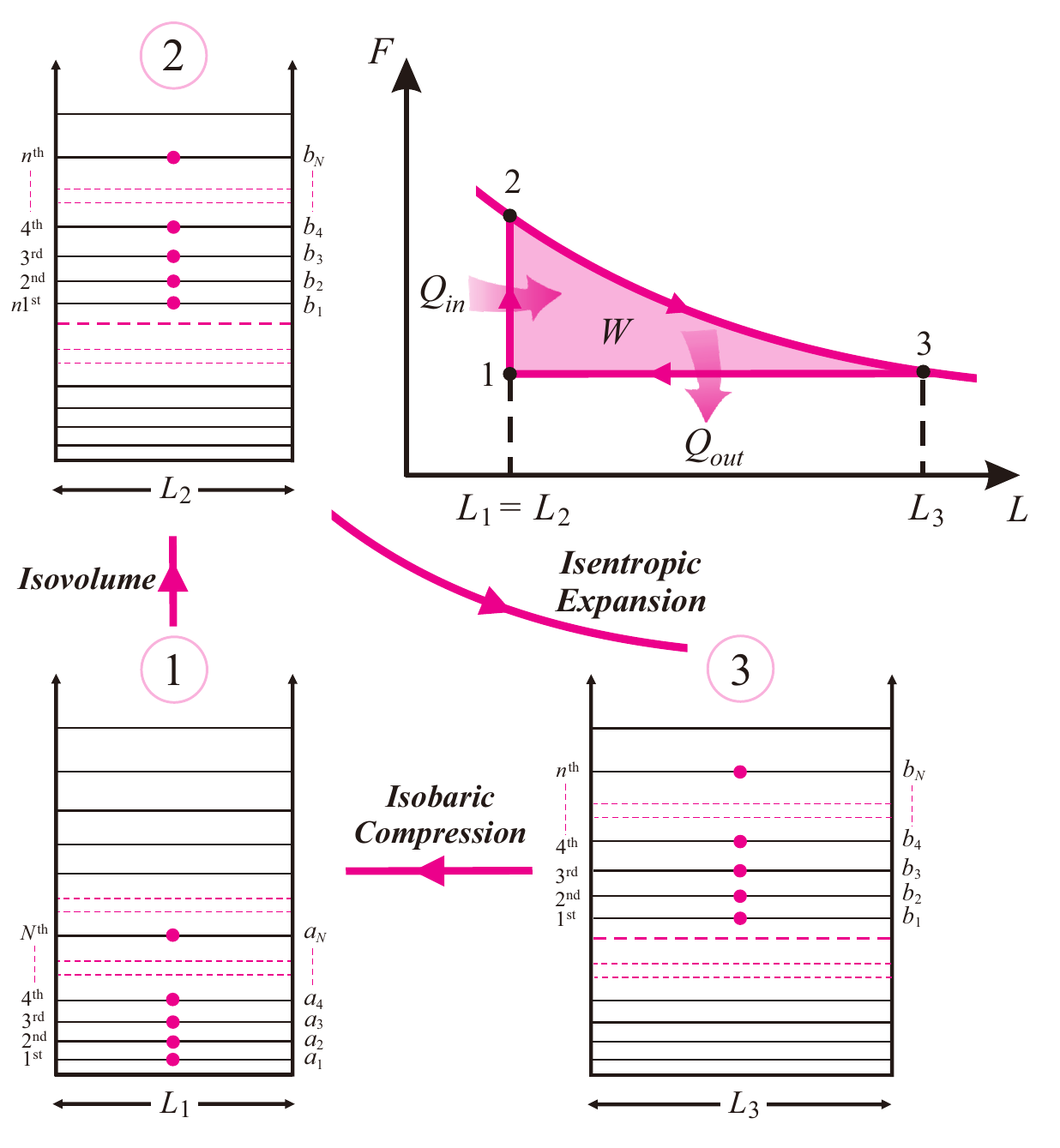}
\caption{\label{fig1} Heat engine based on Lenoir cycle by using multilevel quantum system with N noninteracting fermions as working substance. }
\centering
\end{figure}

Like the classical, quantum Lenoir engine supposed to have three processes within one cycle, as shown in Figure \ref{fig1}. The cycle starts with isovolume process which volume of the well remain the same, $L_2=L_1$. As the result, the change of work of this process will become zero and the internal energy of the engine will only be affected of the heat injected from the hot reservoir, $dU=dQ \thickapprox Q_{in}$.

At the beginning of this process, one of which Fermion particles will nest at ground state and another one is at the first excited level and so on, because these particles should obey Pauli’s exclusion principle \cite{SSingh1.2020,SSingh2.2020,SSingh3.2020}. We call these states as the initial states of the non-interacting fermions, $\sum\limits_{i}^{N}\sum\limits_{n} \left[ a_{n}^{(i)} \right] ^{2}n^{2}=S_{i}$, where $a_{n}^{(i)}$ acts as the index of the fermions at different $n$-level of eigen-energy as seen in Figure \ref{fig1}. So the initial energy of this process can be written as
\begin{align}
E_{1} &= \frac{\pi^{2} \hbar^{2}}{2 m L_{1} ^{2}} S_{i} \label{eq1}
\end{align}
which $E_{1}$ indicates energy of fermions at $L_1$ and the final energy of this process will be 
\begin{align}
E_{2} = \frac{\pi^{2} \hbar^{2}}{2 m L_{1} ^{2}} S_{f} \label{eq2}
\end{align}
as $E_2$ is energy at $L_2$ with final states is $\sum\limits_{i}^{N}\sum\limits_{n}\left[b_{n}^{(i)}\right] ^{2}n^{2}=S_{f}$ and the various $n$-level energy of fermions’ index of these final states showed as $b_{n}^{(i)}$. By subtracting both energy of final and initial states of fermion in this process, we got
\begin{align}
Q_{in} = \frac{\pi^{2} \hbar^{2}}{2 m L_{1} ^{2}} ( S_{f} - S_{i} ) . \label{eq3}
\end{align}

After the engine reaches $L_2$, isentropic process will occur where the system is closed to external energy, $dU = dW$. In this process, the states of fermions will remain constant as the volume slowly expand using the energy gained from the first process, $dW = -dE = \int F dL$. By product of the first process, based on Figure \ref{fig1}, this process will contain the final states of fermions
\begin{align}
F_{23} = \frac{\pi^{2} \hbar^{2}}{m L ^{3}} S_{f} \label{eq4}
\end{align}
where $L$ is volume of engine that moves from $L_2$ to $L_3$ respectively. So the total work in this process would be
\begin{align}
W_{23} = \frac{\pi^{2} \hbar^{2}}{m} S_{f} \int_{L_{2}}^{L{3}} \frac{1}{L ^{3}} dL = \frac{\pi^{2} \hbar^{2}}{2 m L_{1} ^{2}} \left( 1 - \frac {1}{r^{2}} \right) S_{f} \label{eq5}
\end{align}
where $r$ represents the ratio of final and initial volume of the Lenoir cycle, $L_{3}/L_{1} = r$.

Lenoir heat engine will move back to its initial volume in isobaric process as shown in Figure \ref{fig1}. By considering the first law of thermodynamic, we able to derive the heat and work done in the engine, $dU = dQ - dW$.
where $dQ$ is the change of heat along the process and $-dW$ is the output work of the engine, respectively. The force of the well will be kept constant, $dW = -F_{const} \int dL$. Considering the constant force of each initial and final point of this process, we achieve relation between states and volume of the system
\begin{align}
\frac{S_f}{S_i} = \left( \frac{L_3}{L_1} \right) ^{3} = r^3 . \label{eq6}
\end{align}
Furthermore, to derive the work done, we can choose either final or initial force. The total work for isobaric process is
\begin{align}
W_{31} = -F_{const} \int_{L_{3}}^{L{1}} dL = - \frac{\pi^{2} \hbar^{2}}{m L_{1} ^{2}} \left( r - 1 \right) S_{i} \label{eq7}
\end{align}
where the negative sign shows that the total work done goes out of engine as product of cycle.

Whereas the internal energy of the system can be obtained by considering the energy of initial point and final point of the process, written as
\begin{align}
U_{31} = \int_{E_{3}}^{E{1}} dE = \frac{\pi^{2} \hbar^{2}}{2 m L_{1} ^{2}} \left( 1 - r \right) S_{i} . \label{eq8}
\end{align}
Thus, by considering heat exchange of system to surrounding (as the product of cycle), we obtain $Q_{out}$ as
\begin{align}
Q_{out} = \left\lvert\frac{\pi^{2} \hbar^{2}}{2 m L_{1} ^{2}} (3r-3) S_{i}\right\rvert \label{eq9}
\end{align}
where absolute bracket refers only to the magnitude. The “direction” of heat flow is shown by the negative sign that heat flow out of the engine.

\section{Existence of Heat Leakage} \label{sec3}
Leakage is normal phenomenon that occasionally occurs in mechanical machine \cite{FMoukalled.1995}. In this quantum study, heat leakage appears due to interaction between two baths, i.e. heat bath and sink bath that both have high energy degrees and fast relaxation of the particle within engine models \cite{JWang.2012}. Heat leakage appears to both energy bath on quantum engine model as
\begin{align}
Q_{in} = \frac{\pi^{2} \hbar^{2}}{2 m L_{1} ^{2}} ( r^3 - 1) S_{i} + \dot{Q}_{r} \tau ,\label{eq10}
\end{align}
and
\begin{align}
Q_{out} = \left\lvert\frac{\pi^{2} \hbar^{2}}{2 m L_{1} ^{2}} (3r-3) S_{i}\right\rvert  + \dot{Q}_{r} \tau \label{eq11}
\end{align}
where $\dot{Q}_r$ represents heat leakage rate and $\tau$ is total time for one cycle. $\tau$ can be derived by considering the total length change from one cycle as follow
\begin{align}
L_{total} = L_{12} + L_{23} - L_{31} = 2 \left( L_3 - L_1 \right). \label{eq12}
\end{align}
So,
\begin{align}
\tau = L_{total} / \bar{v} = 2 \left( L_3 - L_1 \right) / \bar{v} \label{eq13}
\end{align}
$L_{total}$ is the total volume changes in one cycle and $\bar{v}$ presents average speed for all volume changes. 

\section{Quantum Lenoir Engine Properties as Heat Leakage Considered} \label{sec4}

The efficiency of heat engine shows ratio between input energy and output works of the heat engine in one cycle. This principle can be expressed as
\begin{align}
\eta = \frac{W_{total}}{Q_{in}} \label{eq14}
\end{align}
where $W_{total}$ is the sum of all works done by the engine in one cycle. This total works derived by
\begin{align}
W_{total} = \oint F dL = \frac{\pi^{2} \hbar^{2}}{2 m L_{1} ^{2}} ( r^3 - 3r + 2) S_{i} .\label{eq15}
\end{align}
Therefore, by substituting $W_{total}$ and $Q_{in}$ of Equation (\ref{eq15}) and Equation (\ref{eq10}), respectively, to Equation (\ref{eq14}), we obtained
\begin{align}
\eta = \frac{\frac{\pi^{2} \hbar^{2}}{2 m L_{1} ^{2}} ( r^3 - 3r + 2) S_{i}} {\frac{\pi^{2} \hbar^{2}}{2 m L_{1} ^{2}} ( r^3 - 1) S_{i} + \dot{Q}_{r} \Big( 2(L_3 - L_1) \Big) / \bar{v}} \label{eq16}
\end{align}
where the value of $\dot{Q}_{r}$ assumed as a comparison of the dimension of Equation (\ref{eq16})
\begin{align}
\dot{Q}_{r} = \left( \alpha \frac{S_f}{S_i} \right) \frac{\pi^{2} \hbar^{2} \bar{v}}{2 m L_{1} ^{3}} S_{i} \label{eq17}
\end{align}
$\alpha \left( {S_f}/{S_i} \right)$ is considered to be the heat leakage parameter with ratio of initial and final states. Thus, we got
\begin{align}
\eta = \frac{\eta _{normal}}{1 + \alpha r^3 \frac{2}{3} \eta _{normal}} \label{eq18}
\end{align}
as efficiency of quantum Lenoir heat engine with heat leakage, whereas $\eta _{normal}$ is the efficiency Lenoir heat engine without leakage, $\eta _{normal} = 1 - 3 \left( \frac{r-1}{r^3-1} \right)$.

\begin{figure*}[htp]
\centering
\includegraphics[width=\linewidth,height=6.0cm]{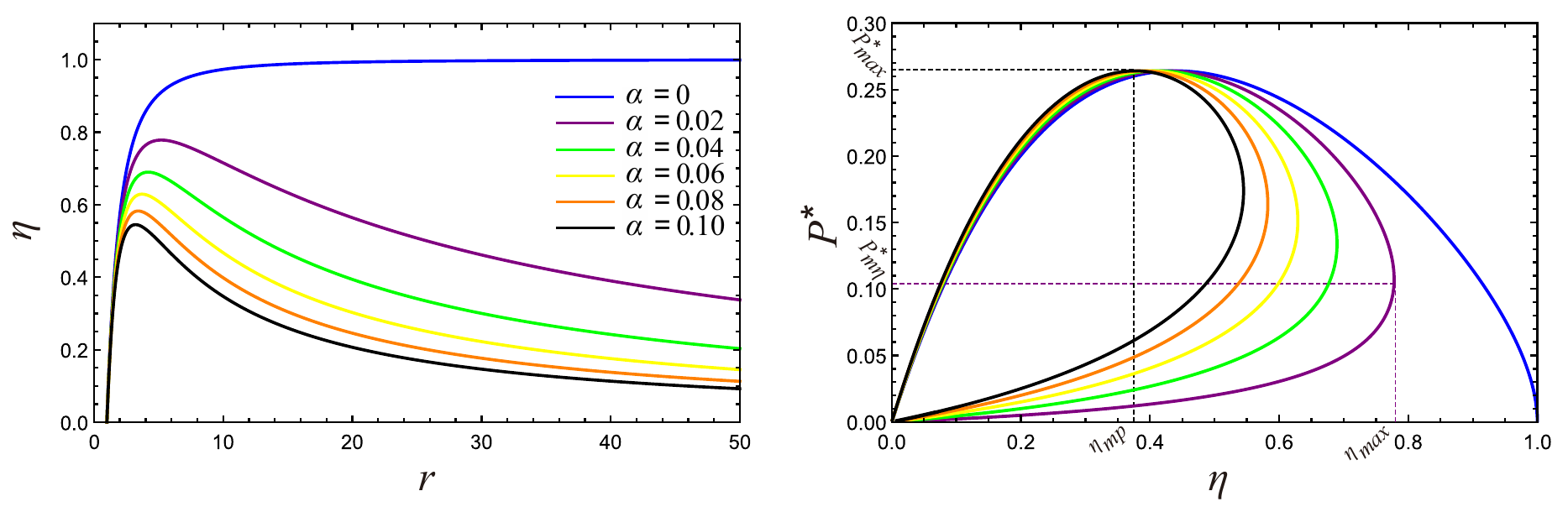}
\caption{\label{fig2} Efficiency vs. compression ratio curves (left) and  dimensionless power vs. efficiency curves (right).}
\end{figure*}

As shown in Figure \ref{fig2}, efficiency of the engine decrease as compression ratio goes higher for each model with leakage. The one without leakage remains steady at high efficiency even the ratio goes up. This proves that the engine efficiency depends on the level of heat leakage that occurs in the engine. The gap between each level of heat leakage become closer. It means a little leakage will affect the engine but gradually vanish as the leakage goes higher. However, at those higher levels of leakage, the engine will not be efficient enough to apply as a daily engine.

We obtained the maximum value of efficiency at the specific point of compression ratios by differentiating Equation (\ref{eq18}) to compression ratio and equalized to zero, $\left( \frac{\partial{\eta}}{\partial{r}} \right) _{r = r_{m \eta}} = 0$. The derivative result is
\begin{align}
\alpha \left( - 2r_{m \eta} ^{4} - 4r_{m \eta} ^{3} + 12r_{m \eta} ^{2} \right) + 6r_{m \eta} + 3 = 0 \label{eq19}
\end{align}
where $r_{m \eta}$ refers to the possible value of $r$ for maximum efficiency, $\eta_{max}$, and the results can be seen in Table \ref{tab1}.

The power output of engine can be derived using the total works done and the time required for a cycle
\begin{align}
P = \frac {W}{\tau} .\label{eq20}
\end{align}
By inputting Equation (\ref{eq15}) and (\ref{eq13}) to Equation (\ref{eq20}), we obtained
\begin{align}
P =\frac{\pi^{2} \hbar^{2}}{4 m L_{1} ^{3}} S_{i} \bar{v} \left( \frac {r^3 - 3r + 2}{r - 1} \right) \label{eq21}
\end{align}
where the power is not explicitly affected by heat leakage. The power needs to be transformed to dimensionless form to find the relation of power and the efficiency of engine written as
\begin{align}
P^* = \left( \frac {r^3 - 3r + 2}{2(r^4 - r^3)} \right) \label{eq22}
\end{align}
where the relations visualize in Figure \ref{fig2}. The curves show that engine with leakage has characteristic to go down to its zero efficiencies as the compression ratio increased by showing loop shapes. As the level of leakage increase, the loop shapes escalate steepened. 

\begin{table}[h]
\begin{ruledtabular}
\caption{Optimizations properties for various level of leakage, $\alpha$}\label{tab1}%
\begin{tabular}{cccc}
$\alpha$    & $r_{m \eta}$   & $\eta_{max}$  & $\eta_{mp}$ \\
\hline
0.00      & $\infty$   & 1          & 0.4398  \\
0.02      & 5.2095     & 0.7781     & 0.4256  \\
0.04      & 4.1964     & 0.6897     & 0.4123  \\
0.06      & 3.7178     & 0.6289     & 0.3998  \\
0.08      & 3.4224     & 0.5825     & 0.3881  \\
0.10      & 3.2163     & 0.5449     & 0.3770  \\
\end{tabular}
\end{ruledtabular}
\end{table}

Every level of leakage (from 0 to 0.10) reaches the maximum value of power but at different points. In the same way as efficiency, maximum value of power can be obtained by differentiating Equation (\ref{eq22})

\begin{figure}[hbp]
\centering
\includegraphics[width=\linewidth]{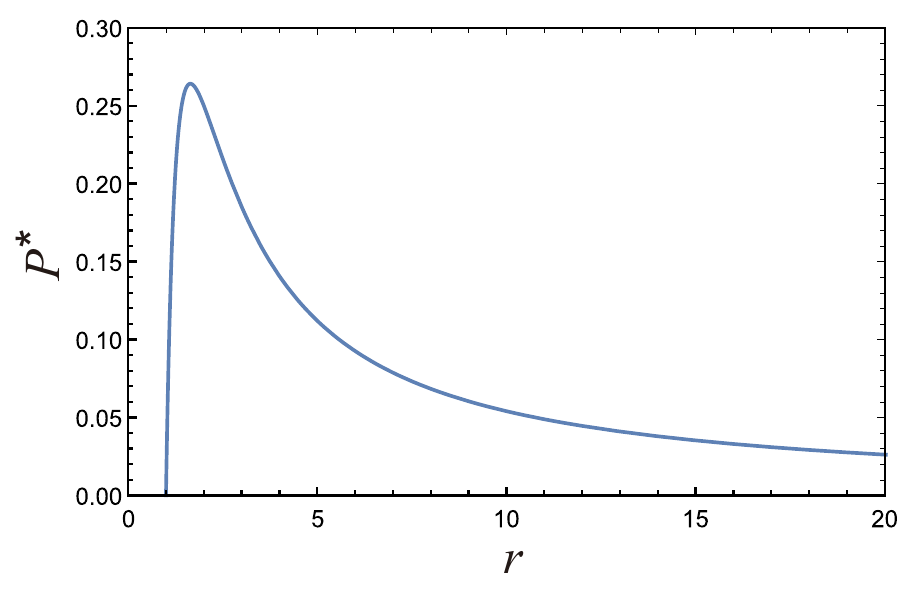}
\caption{\label{fig3} Dimensionless power vs. compression ratio which is not affected by any level of heat leakage, $\alpha$.}
\end{figure}

\begin{align}
\left( \frac{dP^*}{dr} \right) _{r=r_{mp}} = r_{mp}^2 + 2r_{mp} - 6 = 0 \label{eq23}
\end{align}
where $r_{mp}$ represents the compression ratio at maximum power and the result is $r_{mp} = 1.646$ which the maximum power is $P^*_{max} = 0.2641$, as shown in Figure \ref{fig3}. Inserting this ratio to Equation (\ref{eq18}) gives us the value of efficiency at maximum power for each variation of leakage levels, as seen in Table \ref{tab1}.

\section{Clausius Relation of Leaking Engine} \label{sec5}

\begin{figure*}[htp]
\centering
\includegraphics[width=\linewidth]{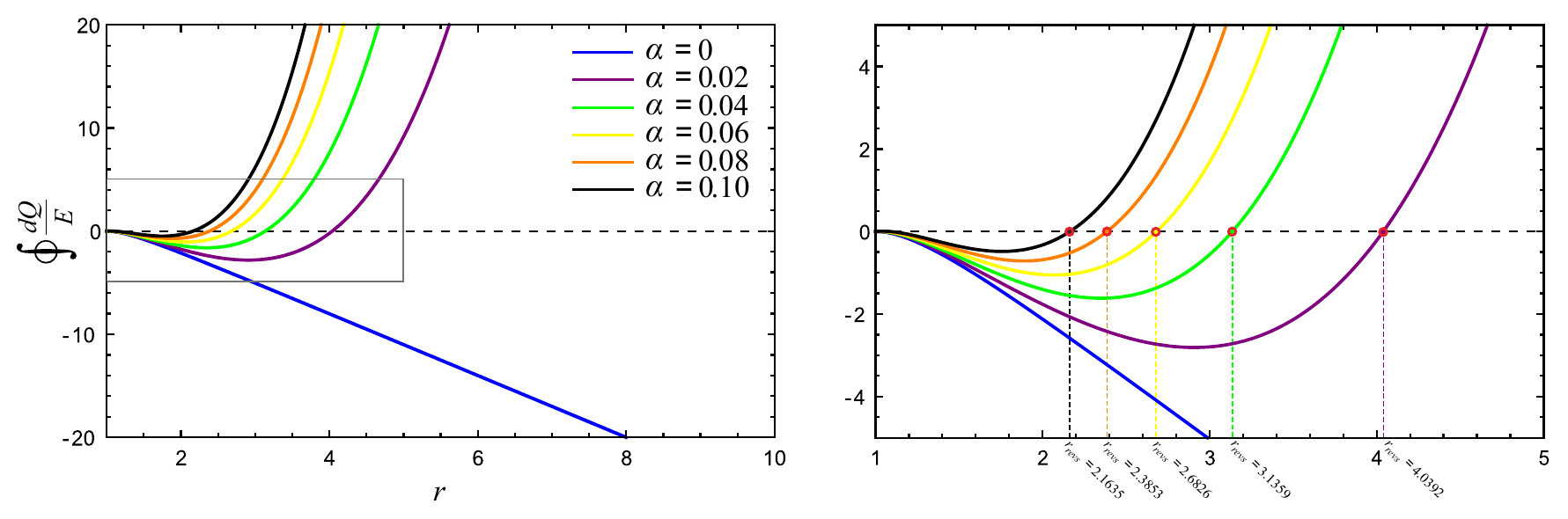}
\caption{\label{fig4} Clausius relation vs. Compression ratio of quantum Lenoir engine model with various heat leakage levels, (right) zoomed version. Red points represent $r_{revs}$ as limitation of compression ratio.}
\end{figure*}

By evaluating the sums transformation of cycling process, Rudolf Clausius stated an inequality to distinguish reversible and irreversible cycles by evaluating the heat flows into engine and the heat flows out. This inequality can be applied to determine whether a model of engine has reversibility or not. Clausius’ inequality written as
\begin{align}
\oint{\frac{dQ}{E}} \leq 0 \label{eq24}
\end{align}
where $dQ$ is the changes of heat within cycle and E is the energy baths of the model. If left-hand side is equal to zero means the cycle is reversible, while inequality represents an irreversible cycle \cite{HSLeff.2018}.

The changes of heat is given from Equation (\ref{eq10}) and (\ref{eq11}). As for the energy baths are 
\begin{align}
E_H = \frac{\pi^{2} \hbar^{2}}{2 m L_{1} ^{2}} S_{f}\label{eq25}
\end{align}
and
\begin{align}
E_C = \frac{\pi^{2} \hbar^{2}}{2 m L_{1} ^{2}} S_{i}\label{eq26}
\end{align}

which $E_H$ represents the high-energy heat bath and the $E_C$ represents the low-energy sink bath. Substituting the changes of heat and energy baths to Clausius’ inequality, we got Clausius relation as
\begin{align}
\oint{\frac{dQ}{E}} &=\frac{Q_{in}}{E_H}+\frac{Q_{out}}{E_C} \nonumber\\ 
&=\left( 4-3r-\frac{1}{r^3} \right) + 2\alpha \left( r^4 - r^3 + r - 1 \right) .\label{eq27}
\end{align}

\begin{table}[t]
\begin{ruledtabular}
\caption{Properties of reversible point on leaking engine}\label{tab2}%
\begin{tabular}{cccc}
$\alpha$    & $r_{revs}$   & $\eta_{revs}$  & $P*_{revs}$\\
\hline
0.00      & $\infty$   & -            & -       \\
0.02      & 4.0392     & 0.7651       & 0.1394  \\
0.04      & 3.1359     & 0.6674       & 0.1779  \\
0.06      & 2.6826     & 0.5971       & 0.2041  \\
0.08      & 2.3853     & 0.5402       & 0.2238  \\
0.10      & 2.1635     & 0.4908       & 0.2392  \\
\end{tabular}
\end{ruledtabular}
\end{table}

Based on Equation (\ref{eq27}), we obtained Clausius relation versus compression ratio curves as in Figure \ref{fig4}. The curves of engine with leakage go across the $x$-axis of $r$ as we raised the compression ratios. This condition approves that the presence of heat leakage within engine will ravage the Clausius inequality, because the heat that supposed to flows out vanished as a leaking heat. Eventually, the system will have an equal amount of heat flows in and out at some exact point of compression ratio, but at high level of leakage, there will be more heat vanished without flowing out through the sink bath or even building up the works.

This groundbreaking result can be explained by looking out at the fundamental of the inequality of the Clausius relation itself. Clausius relation of the heat describes on how much heat that flow into the engine compared to the heat flow out itself by equaling total heat in, $Q_{in}$, divide hot reservoir, $E_{H}$, to total heat out, $Q_{out}$, divide cold reservoir, $E_{C}$. For an irreversible engine, the total heat out divide its reservoir will be larger nominally than the heat and reservoir at the entrance side. 

Lenoir engine should be an irreversible engine if this theorem applied to it, but the presence of heat leakage altered the total heat in and out of the system. While the wall of the quantum heat engine expanded and compressed, some of the engine that has been infused will be vanished as leakage. As the heat flows out, the amount has been dramatically decreased. As result, the equalization of the both side of reservoir will describe the engine as reversible engine. So, this "reversible" norms actually representation of the equal amount of heat that depleted by the heat leakage occured in the engine.

To keep the model intact to standard engine, there should be some limitations to compression ratio of engine can be made. This limitation is obtained by finding the roots of the curves as in Table \ref{tab2}. Aside from these limitations, eventually, we found one reversible point for each model with various leakage within, written as $r_{revs}$. By substituting these $r_{revs}$ to the model optimization properties, we got the reversible efficiency, $\eta_{revs}$, and the reversible power, $P^*_{revs}$, as shown in Table \ref{tab2}. The engine efficiency decreased as we raised the variations of heat leakage (which means compression ratio also increased), in other hand, the power output of the engine model leveled up because the value of the power do not affected by the value of parameter heat leakage.

 \section{Conclusion} \label{sec6}
 We successfully constructed the Heat Engine model based on Lenoir Cycle by using multilevel quantum system with N noninteracting fermions as a working substance. We obtain that the efficiency of the engine depends on the compression ratio and also leakage parameter, whereas the efficiency is independent of the number of particles. Considering heat leakage as a physics parameter gives a different view of quantum heat engines modeling, especially Lenoir cyclic engine. The loop curves strengthen that the efficiency of heat engine is explicitly affected by the value of leakage parameter $\alpha$, but otherwise for the power output of the engine. The increasing level of leakage shift the parabolic-like curves become the steepened loop. Moreover, we have also derived the optimized properties of the engine model with and without leakage. The reversible model with a specific compression ratio was acquired while we derived the Clausius relation of this model. This new strike could be a new field study and more evaluation on particle itself, especially Fermions, and investigation on multiple cycle this reversible model would significantly improve the model's optimizations and evaluations.

\begin{acknowledgments}
 This work was financially supported by Andalas University with research grant No. T/38/UN.16.17/PT.01.03-IS-RDP/2021.
\end{acknowledgments}



\bibliography{apssamp}

\end{document}